\documentclass[
reprint,
superscriptaddress,
amsmath,amssymb,
aps,
floatfix,
]{revtex4-1}

\usepackage{graphicx}
\usepackage{dcolumn}
\usepackage{bm}
\usepackage{braket}
\usepackage[german,english]{babel}
\usepackage[colorlinks=true,linkcolor=blue,citecolor=blue,breaklinks]{hyperref}
\usepackage[pdftex]{color}
\usepackage{ulem}
\usepackage{comment}
\usepackage{natbib}
\usepackage{xcolor}

\begin{document}
	
	\title{Fractional corner charges in a 2D super-lattice Bose-Hubbard model}
	
	\author{Julian Bibo}
	\affiliation{Department of Physics, T42, Technical University of Munich, D-85748 Garching, Germany }
	\affiliation{Munich Center for Quantum Science and Technology (MQCST), Schellingstr. 4, D-80799 Munich, Germany }
	\author{Izabella Lovas}
	\affiliation{Department of Physics, T42, Technical University of Munich, D-85748 Garching, Germany }
	\affiliation{Munich Center for Quantum Science and Technology (MQCST), Schellingstr. 4, D-80799 Munich, Germany }
	\author{Yizhi You}
	\affiliation{Princeton Center for Theoretical Science, Princeton University, NJ, 08544, USA}
	\author{Fabian Grusdt}
	\affiliation{Munich Center for Quantum Science and Technology (MQCST), Schellingstr. 4, D-80799 Munich, Germany }
	\affiliation{Department of Physics and Institute for Advanced Study,Technical University of Munich, D-85748 Garching, Germany }
	\affiliation{Department of Physics and Arnold Sommerfeld Center for Theoretical Physics (ASC), Ludwig-Maximilians-Universit\"at M\"unchen, Theresienstr. 37, D-80333 Munich, Germany} 
	\author{Frank Pollmann}
	\affiliation{Department of Physics, T42, Technical University of Munich, D-85748 Garching, Germany }
	\affiliation{Munich Center for Quantum Science and Technology (MQCST), Schellingstr. 4, D-80799 Munich, Germany }		
	\date{\today}
	
	\begin{abstract}
		We study a two dimensional super-lattice Bose-Hubbard model with alternating hoppings in the limit of strong on-site interactions. We evaluate the phase diagram of the model around half-filling using the density matrix renormalization group method and find two gapped phases separated by a gapless superfluid region. We demonstrate that the gapped states realize two distinct higher order symmetry protected topological phases, which are protected by a combination of charge conservation and $C_4$ lattice symmetry. The phases are distinguished in terms of a quantized fractional corner charge and a many-body topological invariant that is robust against arbitrary, symmetry preserving edge manipulations. We support our claims by numerically studying the full counting statistics of the corner charge, finding a sharp distribution peaked around the quantized values. These results are experimentally observable in ultracold atomic settings using state of the art quantum gas microscopy.
	\end{abstract}
	
	\maketitle
\section{Introduction}
The role of symmetries in topological properties of strongly correlated many-body systems currently receives a lot of attention.
There has been an intense effort to classify symmetry protected topological phases of matter, which are characterized by their edge modes and topological invariants~\citep{Schuch2011,Pollmann2010,Chen2013a,Song2017,Kitaev2001,Ryu2002,Delplace2011,Grusdt2013,Chen2011,Peng2017}. 
Recent studies have refined the classification to also include crystalline symmetries, leading to protected gapless corners or hinge states. 
These have been extensively explored in non-interacting fermionic systems, referred to as higher order topological~(HOTI) phases~\citep{Benalcazar2017,Benalcazar2017a,Langbehn2017,Schindler2018,Schindler2019b,Blanco2019,Bouhon2019} or fragile phases \cite{po2018fragile,Cano2018,else2019fragile}, and interacting higher order symmetry protected topological~(HOSPT) phases~\citep{isobe2015theory,song2017interaction,Dubinkin2018,You2018,thorngren2018gauging,else2019crystalline,SongHermele2017,Rasmussen2018}.
Interacting bosonic HOSPT phases are partially classified in terms of group cohomology based on the interplay of these global and crystalline symmetries \cite{thorngren2018gauging,Rasmussen2018}.
Despite the rapid progress in the theoretical understanding of HOSPT phases, the experimental observation and proposal of interacting HOSPT states remains challenging, since most of the theoretically studied HOSPT models involve special plaquette interactions between spins, or require artificial gauge potentials. 
%
In spite of these difficulties, the rapid development in experimental techniques in ultracold atomic settings opened up unprecedented possibilities to study topological phases~\citep{Bloch2008,Ruostekoski2002,Tarruell2012,Atala2013,Aidelsburger2014,Lohse2015,Tai2017,Miyake2013,Aidelsburger2013,Struck2012}. 
In particular, the site resolved control of ultracold atoms in optical lattices offers an ideal opportunity to study corner modes and their deep connection to the topological properties of the bulk in the presence of interactions. 

In this work, we propose an experimentally accessible two dimensional~(2D) ultracold atomic system, a super-lattice Bose-Hubbard model~(SL-BHM), supporting an HOSPT phase protected by a combination of a $U(1)$ charge conservation and $C_4$ lattice symmetry. 
Although the $U(1) \times C_4$ alone does not protect gapless corner modes \cite{You2019a}, we show that our boson model still exhibits symmetry anomalies at the corners in the form of fractional charges that characterize the phase. 
Namely, the fractionally charged corner cannot be annihilated without a bulk gap closing. 
We discuss a many-body invariant that goes hand in hand with a quantized fractional charge localized around the corners of a system with sharp edges. 
By numerically studying the full counting statistics of the corner charge, we show that the system exhibits sharp distributions peaked around the quantized values. 
%

\begin{figure}[t!]
\includegraphics[width=0.48\textwidth]{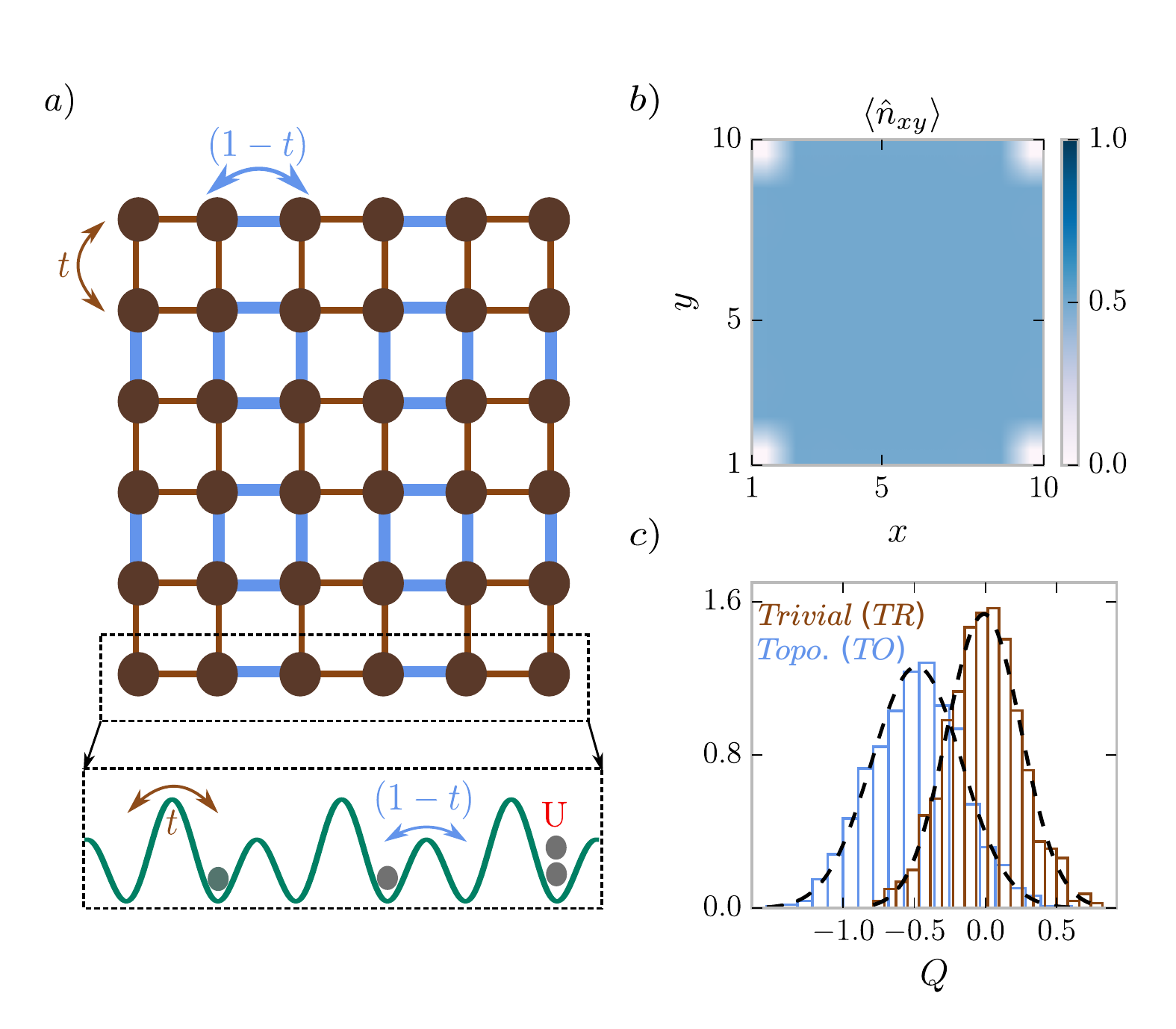}
\caption{2D SL-BHM showing an HOSPT phase with quantized fractional corner charges. \textbf{a)} The SL-BHM has a $2\times 2$ unit cell, with two different hopping amplitudes inside and between unit cells, $t$ and $1-t$, respectively. The ground state is topological trivial~(non-trivial) for $t=1{\,}(t=0)$. \textbf{b)} Average occupation number of lattice sites in the topological phase with particle number $N=L^2/2-2$, displaying four holes localized around the corners, giving rise to fractional charges $Q_{\rm corner}=1/2$. \textbf{c)} The full distribution function of corner charges for the trivial~(topological) phase, peaked around a quantized fractional part $0{\,}(1/2)$. We used the parameters $N=L^2/2{\,}(L^2/2-2)$, $t=0.9{\,}(0.1)$, $U=32$ and $\xi_{\rm env}=3.2{\,}(3.08)$.} 
\label{fig:Lattice}
\end{figure}

\section{Model} We consider the following SL-BHM on a two dimensional square lattice of size $L\times L$ with $L$ even,
	\begin{align}
		\label{Ham.}
		\hat{H}=&-\left[\sum^{L-1}_{x=1}\sum^{L}_{y=1}\left(t(x)\hat{a}^{\dagger}_{x,y}\hat{a}_{x+1,y}+\text{h.c.}\right) +x\leftrightarrow y\right] \nonumber\\
		&+\frac{U}{2}\sum_{x,y=1}^L\hat{n}_{x,y}\left(\hat{n}_{x,y}-1\right),
	\end{align}where $\hat{a}^{\dagger}_{x,y}$ $\left(\hat{a}_{x,y}\right)$ is the creation~(annihilation) operator at site $(x,y)$, and $\hat{n}_{x,y}=\hat{a}^{\dagger}_{x,y}\hat{a}_{x,y}$. The particles can tunnel between neighboring sites with modulated hopping amplitudes $t(\zeta),{\,}\zeta\in\{x,y\}$, 
	\begin{equation}
	t(\zeta)=\begin{cases}
	t, & {\rm for} \;\;\zeta\in\{1,3,...,L-1\}\\
	1 - t, &{\rm for} \;\;\zeta\in\{2,4,...,L-2\}
	\end{cases}
	\end{equation} where $t\in[0,1]$, while the parameter $U\geq{0}$ controls the on-site repulsion between the particles (see~Fig.~\ref{fig:Lattice}a). 
We study Hamiltonian~\eqref{Ham.} at a fixed number of particles and open boundary conditions, focusing on two particle number sectors around half-filling, $N=L^2/2\equiv N_0$ and $N=N_0-2$. 
This model is the bosonic counterpart of the 2D free fermion Benalcazar-Bernevig-Hughes~(BBH) model, a well known HOTI hosting gapless corner states~\citep{Benalcazar2017a}. 
However, let us emphasize that in terms of experimental realization the SL-BHM has an enormous advantage over the BBH model. While the BBH model requires a magnetic flux $\pi$ inserted through each plaquette to ensure a bulk gap, the 2D SL-BHM considered here does not involve a magnetic field, making it more easily accessible in experiments. 
Note that the 2D SL-BHM model and the BBH model are equivalent if both $U\to\infty$ and $t=0{\,}(t=1)$, while either of those conditions alone is insufficient~(see Appendix). 

\begin{figure}[t]
\includegraphics[width=0.48\textwidth]{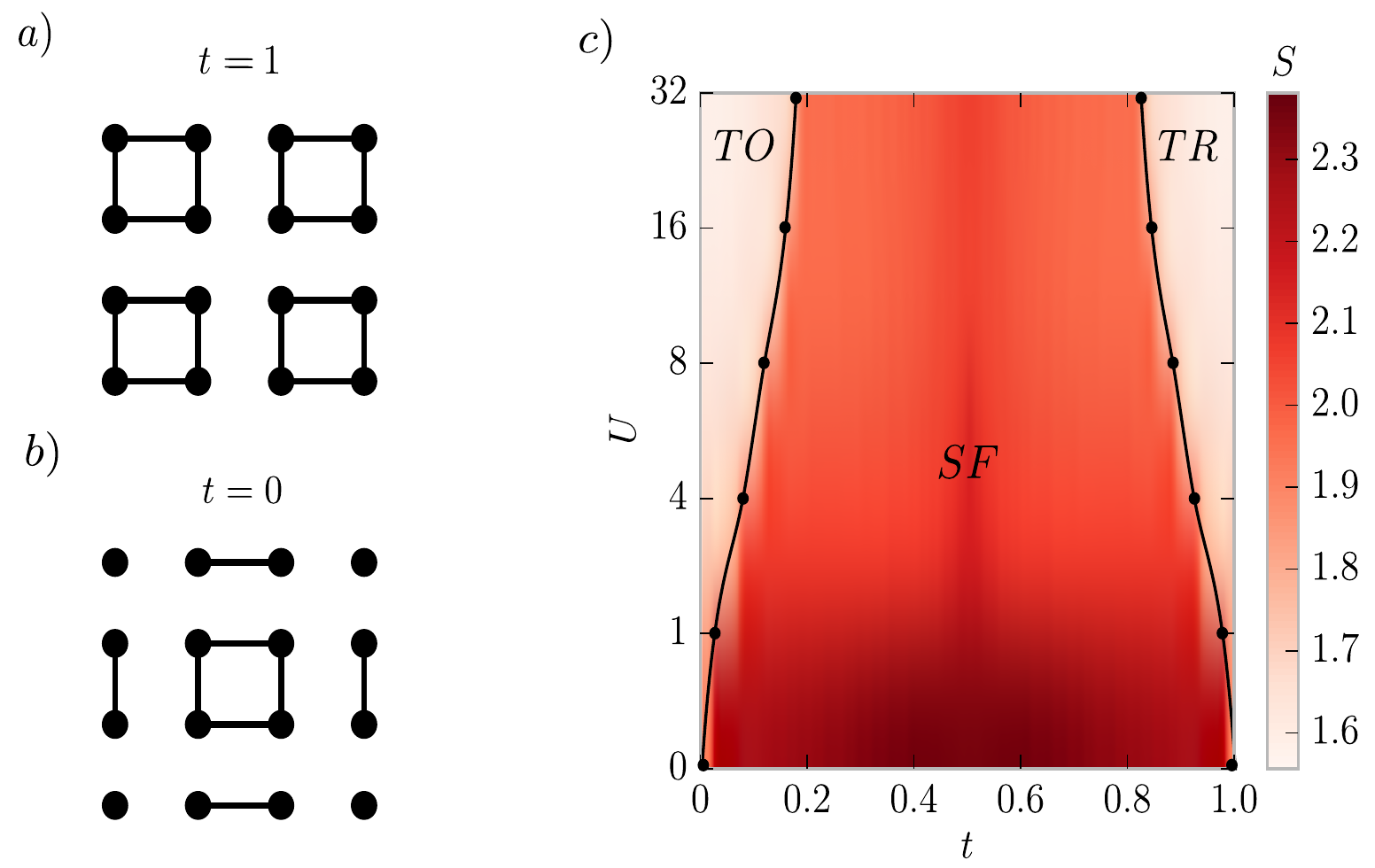}
\caption{Exactly solvable limits of both phases are shown for the trivial (TR) and topological phase (TO) in \textbf{a)} and \textbf{b)}, respectively. In the topological phase, the corner of the lattice is completely decoupled from the rest of system. Panel \textbf{c}) shows the bipartite entanglement entropy $S$ against the interpolation parameter $t$ obtained by DMRG on a infinite cylinder with circumference $L_y=6$. The large entanglement is characteristic for the superfluid (SF) phase. The Hilbert space was truncated to maximal four bosons per site and we used kept $\chi=500$ states for the simulations.} 
\label{fig:Phasediagram}
\end{figure}
Hamiltonian~\eqref{Ham.} is $C_{4}$ symmetric with respect to the center of the lattice and preserves the total particle number $\hat N=\sum_{x,y}\hat{n}_{x,y}$. 
For our purposes $U(1)\times C_4$ is the only relevant symmetry.
We focus mainly on a parameter regime of $(t,U)$ where the bulk is incompressible and the average filling in the bulk is $n_0 = 1/2$. 
Before proceeding, let us discuss the exactly solvable limits $t=1$ and $t=0$ to illustrate the emerging corner charges~(in presence of open boundary conditions). 
In both cases there is a finite bulk gap; for strong interactions $U\gg \text{max}(t,1-t)$ the bulk gap is of order $\Delta_{b}\sim \mathcal{O}(\text{max}(t,1-t))$. 
In the \textit{trivial phase}~(TR) with $t=1$ the ground state at half-filling is unique and $C_4$ symmetric~(see~Fig.~\ref{fig:Phasediagram}a). 
In the \textit{topological phase}~(TO) with $t=0$, the four corners are decoupled and each of them can be either filled or empty (see~Fig.~\ref{fig:Phasediagram}b).
To obtain a unique, $C_4$ symmetric ground state with bulk and edges at half-filling, the total particle number has to deviate from exact half-filling $N=N_0$, as $N=N_0\pm 2$.
This filling anomaly~\footnote{The filling anomaly is a consequence of open boundary conditions. For periodic boundary conditions both phases have a unique, $C_4$ symmetric ground state at the same filling $n_0$.}, giving rise to quantized fractional corner charges~(see Fig.~\ref{fig:Lattice}b), was already discussed in the context of HOTI~\citep{Benalcazar2018}. 
Comparing the charge distributions of the two phases, we find a fractional charge $Q_{\rm corner}=1/2$ localized around the corners in the topological phase, measured with respect to the average bulk filling $n_0$~(for a rigorous definition of $Q_{\rm corner}$ see Eq.~\eqref{eq:Q} below).
The phase diagram of the model at half-filling in Fig.~\ref{fig:Phasediagram}c, obtained using the iDMRG method~\citep{White1992,Liang94,Mcculloch2008,Hauschild2018} on an infinite cylinder, shows that the exactly solvable points extend to two gapped phases~(TR and TO, respectively), which are are separated by a gapless superfluid~(SF) regime. 

\section{Topologically distinct HOSPT phases} 
\begin{figure}[t!]
\includegraphics[width=0.48\textwidth]{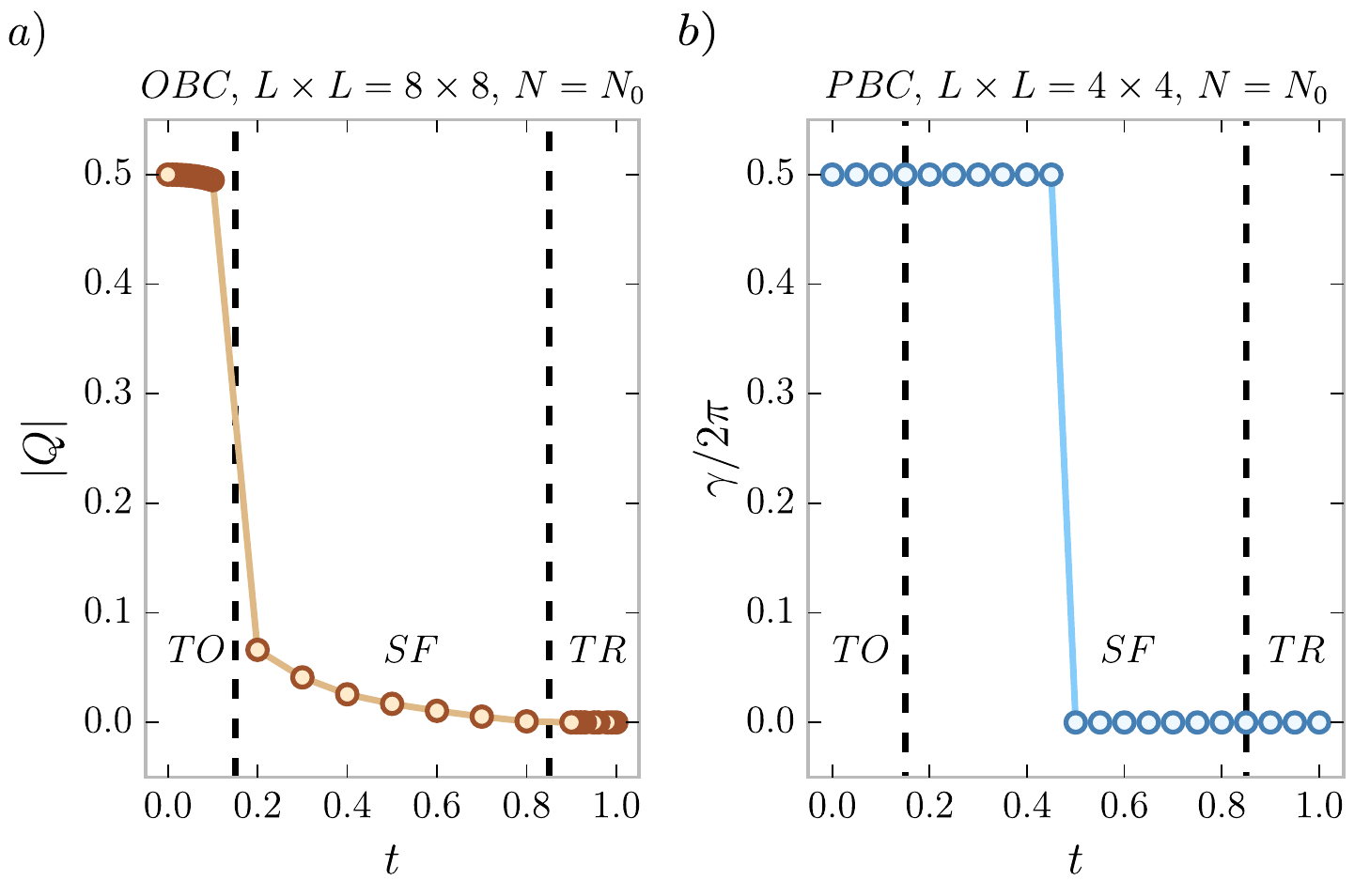}
\caption{\textbf{a)} Corner charge $Q_{\rm corner}$ as a function of hopping $t$ at half-filling $N=N_0$, quantized to $Q_{\rm corner}=0{\,}(1/2)$ in the gapped trivial~(topological) phase. We used DMRG on an $8\times 8$ square lattice, taking $U=32,{\,}\xi_{\rm env} = 2.8$. \textbf{b)} Berry phase $\gamma$ as a function of the hopping parameter, quantized around $\gamma = 0{\,}(\pi)$ in the gapped trivial~(topological) phase. Result is obtained for a $4\times 4$ square lattice with periodic boundary conditions at half-filling.}
\label{fig:Corner}
\end{figure}

Before turning to the detailed study of the topological phase of the 2D SL-BHM, let us briefly review the classification of topological phases. 
Generally, two gapped states belong to the same phase if and only if there is an adiabatic path $\Gamma_{\vec{\lambda}}$ with $\vec{\lambda}\in\Lambda$ in the parameter space $\Lambda$, together with a family of local, gapped Hamiltonians $\{H(\vec{\lambda})\}$, such that the ground states of these Hamiltonians connect the two states without closing the bulk gap $\Delta_{b,\Gamma}({\vec{\lambda}})$ along the path~\citep{Chen2010,Schuch2011}. 
If we additionally impose certain symmetries on the system, we allow only for those paths in $\Lambda$ that explicitly conserve the symmetry.
Such symmetry constraints can significantly enrich the phase diagram.
%
%
Let us emphasize here that HOSPT phases should be robust against closing the \textit{edge} gap $\Delta_e$ of the system~--~two different HOSPT phases can only be connected by closing the \textit{bulk} gap $\Delta_b$.
For the 2D SL-BHM considered here, the required protecting symmetry is a combination of the $C_4$ lattice symmetry and the global $U(1)$ symmetry associated with particle number conservation. 
In presence of these symmetries, the two exactly solvable limits depicted in Figs.~\ref{fig:Phasediagram}a and \ref{fig:Phasediagram}b belong to distinct HOSPT phases. 
Namely, the two phases are labelled by a corner charge, quantized as $Q_{\rm corner}=0\,(1/2){\,}\text{mod}{\,}1$, corresponding to the TR~(TO) phase, that cannot change unless the bulk gap closes.
To demonstrate this, we define polarizations denoted by $P^{\rm edge}$ for each edge. 
Concentrating on a single corner, the change of the corner charge $\Delta Q_{\rm corner}$ due to arbitrary edge manipulations is directly related to the change of the polarizations of the two edges meeting at this corner, $\Delta P^{\rm edge}_{x}$ and $\Delta P^{\rm edge}_{y}$, by the King-Smith-Vanderbilt relation~\cite{King-Smith1993} 
\begin{equation}
\label{KSV}
\Delta Q_{\rm corner} = \sum_{\zeta\in\{x,y\}}{\Delta P^{\rm edge}_{\zeta}}{\,}\text{mod}{\,}1.
\end{equation} 
Due to $C_4$ symmetry, the edge polarizations $\Delta P^{\rm edge}_{x}$ and $\Delta P^{\rm edge}_{y}$ cancel each other, confirming that $Q_{\rm corner}$ is robust against edge manipulations, even if the edge gap closes and thus reflects the properties of the bulk. 
We demonstrate the robustness of quantization numerically in Fig.~\ref{fig:Corner}a, displaying $Q_{\rm corner}$ as a function of $t$, tuned across the TO, SF and TR phases~\footnote{To get a unique ground state in the topological phase we weakly break the $C_4$ symmetry at the corners by a local chemical potential.}. 
We observe two plateaus at $Q_{\rm corner}=0$ and $Q_{\rm corner}=0.5$ in the gapped, incompressible TR and TO phases, respectively.

In addition, a recent work by Araki~et.~al. proposed a Berry phase $\gamma$ as a label for HOSPT phases~\citep{Araki2019}. 
For the 2D SL-BHM, this Berry phase is quantized to $\gamma=0{\,}(\pi){\,}\text{mod}{\,}2\pi$, with the quantization relying on the global $U(1)$ and spatial $C_4$ symmetry.
This quantity distinguishes the two phases as shown in Fig.~\ref{fig:Corner}b~--~further supporting our claim that the 2D SL-BHM realizes two inequivalent HOSPT phases.
Note that the quantized plateaus of the Berry phase in Fig.~\ref{fig:Corner}b persist in the SF phase purely due to finite size effects.
Analogously to the corner charge, the Berry phase is only well-defined in the gapped phases. 
Although, to our knowledge no rigorous proof connecting $\gamma$ and $Q_{\rm corner}$ has been given yet, there is a strong indication that the two different labels provide another example for the bulk-boundary correspondence, with the bulk invariant $\gamma$ going hand in hand with a corner state manifesting in $Q_{\rm corner}$.

\section{Measuring the corner charge} Let us now return to the more detailed discussion of the corner charge and how to measure it. 
We define the operator $\hat{Q}_{\rm corner}$ as
\begin{equation}\label{eq:Q}
\hat{Q}_{\rm corner}=\sum_{x,y}e^{-\vec{r}^2/\xi_{\rm env}^2}\left(\hat{n}_{x,y}-n_0\right),
\end{equation}where $\vec{r}$ gives the position on the lattice and $n_0\equiv 1/2$ denotes the average occupation number.
The width of the Gaussian envelope function is determined by $\xi_{\rm env}$, which has to be much larger than the bulk correlation length $\xi_{\rm env}\gg\xi_{\rm corr}$ and much smaller than the system size $\xi_{\rm env}\ll L$. 
Moreover, in the TO phase there is another relevant length scale, the localization length of a particle or hole state pinned at the corner, $\xi_{\rm p,h}$, and we also demand $\xi_{\rm env}\gg \xi_{\rm p,h}$~\footnote{Within a region of size $\xi^2_{\rm corr}$ or $\xi^2_{\rm p,h}$ particle number fluctuations can be of order $\mathcal{O}(1)$ and therefore destroy the quantization of $Q_{\rm corner}$.}. 
The localization length is defined through the participation ratio~\cite{Murphy2011} $\xi^{2}_{\rm p,h}=\left(\sum_{x,y}\Delta n_{x,y}\right)^2/\sum_{x,y}(\Delta n_{x,y})^2$, where $\Delta n_{x,y} = |\langle{\hat n}_{x,y}\rangle-n_0|\times{\Theta\left(\pm(\langle{\hat n}_{x,y}\rangle-n_0)\right)}$, with ``+'' for $\xi_{\rm p}$ and ``-'' for $\xi_{\rm h}$, and with $\Theta$ denoting the Heaviside step function. For sufficiently large system sizes $\xi_{\rm env}$ becomes independent of $L$, confirming the presence of a localized corner charge.
As argued above, $Q_{\rm corner}=\langle\hat{Q}_{\rm corner}\rangle{\,}\text{mod}{\,}1$ is a good label for the HOSPT phases of the SL-BHM model, quantized to the discrete values $Q_{\rm corner}=0{\,}(1/2)$ in the TR~(TO) phase. 
In fact, the quantum fluctuations of the operator $\hat{Q}_{\rm corner}$ get suppressed for increasing system size, rendering $Q_{\rm corner}$ a well defined quantum number in the thermodynamic limit.

\begin{figure}[t!]
\includegraphics[width=0.48\textwidth]{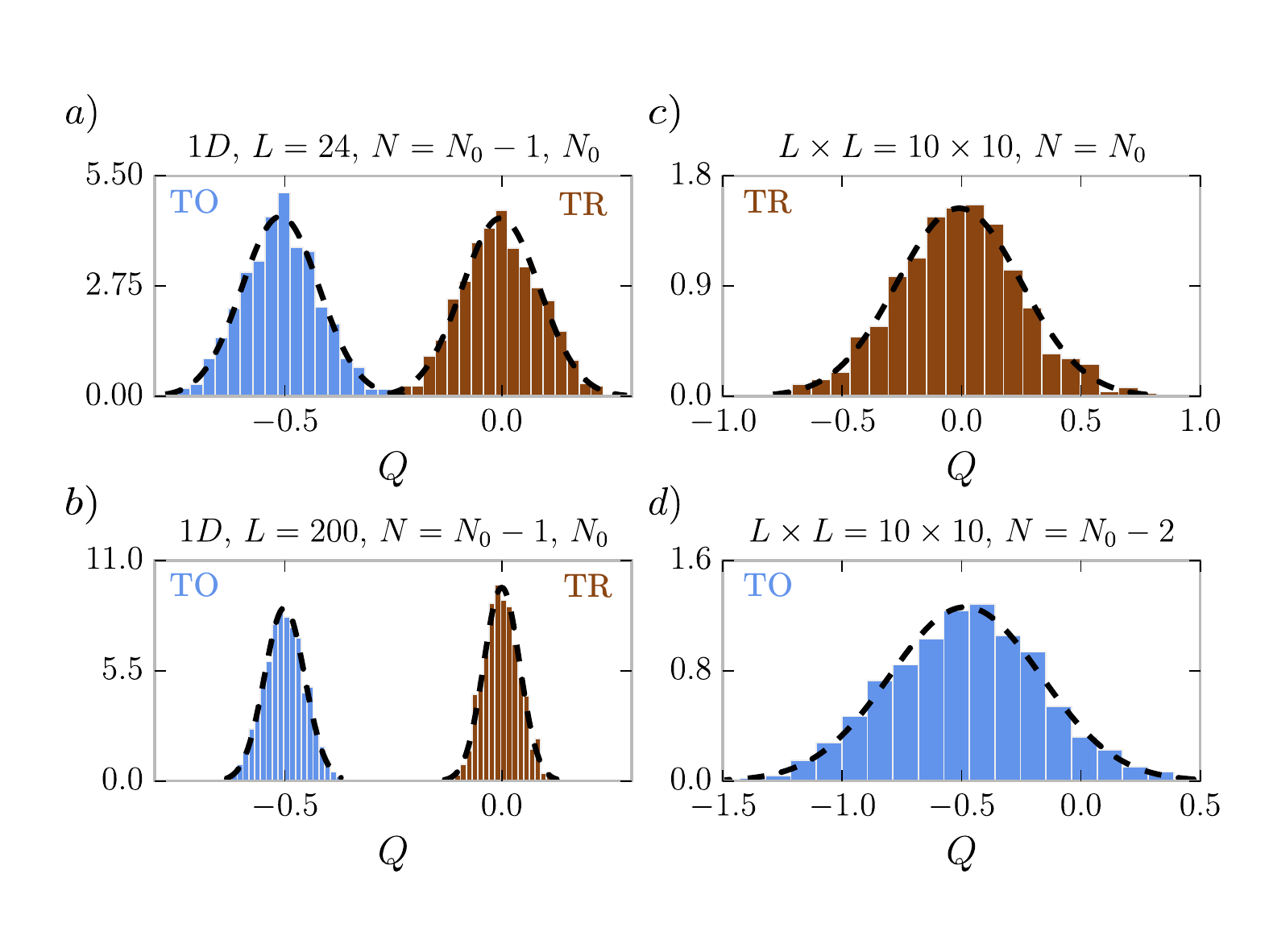}
\caption{Full counting statistics of edge and corner charges. \textbf{a)} and \textbf{b)} FCS of fractional edge charge $Q_{\rm edge}$ of the 1D SL-BHM (defined in~Ref.~\citep{Grusdt2013}) in the TR and TO phases, for two different system sizes $L=24$ and $L=200$. The distribution is peaked around the quantized value $0{\,}(1/2)$ in the TR~(TO) phase and gets sharper with increasing $L$, confirming that $Q_{\rm edge}$ is a good quantum number in the thermodynamic limit. Parameters for the TR~(TO) phase: $N=N_0{\,}(N_0-1)$, $t_1=0.2{\,}(1)$, $t_2=1{\,}(0.2)$, $U=10$. The envelopes are $\xi_{\rm env}=12{\,}(12)$ for $L=24$ and $\xi_{\rm env}=53{\,}(45)$ for $L=200$ respectively. \textbf{c)} and \textbf{d)} FCS of fractional corner charge $Q_{\rm corner}$, measured in the TR~(TO) phase of the 2D SL-BHM for system size $10\times{10}$, peaked around $0{\,}(1/2)$. The width of the distribution should approach to zero in the thermodynamic limit, similar to the 1D case. Results were obtained for $N=N_0{\,}(N_0-2)$, $t=0.9{\,}(0.1)$, $U=32$ and $\xi_{\rm env}=3.2{\,}(3.08)$.}
\label{fig:Statistics}
\end{figure}

Let us note that the gapped phases of the 1D SL-BHM with a bulk at half-filling~\footnote{The 1D SL-BHM also shows a filling anomaly; for the same chemical potential the two phases differ in their total particle number, with $N=N_0{\,}(N_0-1)$ in the TR~(TO) phase.~(See Fig.~1 of Ref.~\citep{Grusdt2013}).} also realize two distinct topological phases, labelled by the bulk polarization $P_{\rm 1D, bulk}$, proven to be quantized to $P_{\rm 1D, bulk}=0{\,}(1/2)$ for a system with periodic boundary conditions~\citep{Grusdt2013}.
This quantization relies on the inversion symmetry only and $P_{\rm 1D, bulk}$ is directly related to a Berry phase picked up by the interacting many-body wave function for twisted boundary conditions~\citep{Hatsugai2006}. 
In a system with sharp edges, bulk-boundary correspondence manifests in a charge localized around the edge, $Q_{\rm 1D,edge}$, such that the changes of $Q_{\rm 1D,edge}$ are related to the changes of bulk polarization through $\Delta Q_{\rm 1D,edge} = \Delta P_{\rm 1D,bulk}$.
Since the latter is quantized and directly related to a topological invariant, the quantum fluctuations of $\Delta Q_{\rm 1D,edge}$ have to vanish in the thermodynamic limit $L\to \infty$ (see~Figs.~\ref{fig:Statistics}a~and~\ref{fig:Statistics}b). 
Though a rigorous analogous argument in 2D is still lacking~\citep{Ono2019}, we present numerical evidence that the fluctuations of the corner charge get suppressed as $L\rightarrow\infty$.
To this end we evaluated the full counting statistics (FCS) of the edge and corner charges of the 1D and 2D SL-BHM, respectively, both in the trivial and in the topological phase. We obtained the ground state of the system using DMRG and then generated single snapshots according to the probability distribution given by the wave function using perfect sampling~\citep{Ferris2012}. 
Figs.~\ref{fig:Statistics}a and \ref{fig:Statistics}b display the FCS of the 1D system in both phases for two different system sizes $L$, clearly demonstrating that the distributions centered around $Q_{\rm 1D, edge}=0{\,}(1/2)$ get sharper as $L$ increases. Figs.~\ref{fig:Statistics}c and \ref{fig:Statistics}d show the 2D FCS in the TR and TO phases, respectively. 
Here the accessible system size $L$ is more limited and the FCS still shows significant finite size effects, in contrast to the 1D case. This results in a broader distribution that is, however, clearly centered around $Q_{\rm corner} = 0{\,}(1/2)$.
As a final remark, let us mention that the energy of the particle state at the corner in the TO phase touches the bulk band before the SF transition; at this point the particle state becomes unstable and melts into the bulk. 
Interestingly, in this region we still find a quantized localized hole state in the vicinity of the corners, similar to the 1D SL-BHM, where a stable quantized hole state was observed at the edge even after the particle already disappeared into the bulk~\citep{Grusdt2013}. 
Importantly, this melting of the particle state is a consequence of on-site repulsion $U$, with no analogue in non-interacting HOSPT phases such as the fermionic BBH model.

\section{Discussion} In summary, we have proposed an experimentally accessible ultracold atomic system, a 2D SL-BHM around half filling, with alternating hoppings $t$ and $1-t$ realizing an interacting HOSPT phase protected by charge conservation and $C_4$ lattice symmetry. 
Relying on DMRG simulations, we explored the phase diagram of the model, and have shown that it hosts two gapped topological phases, separated by a gapless superfluid region. 
Concentrating on the gapped phases, we have argued that they are topologically distinct and differ in terms of a quantized fractional charge localized around the corners, intimately connected to a quantized Berry phase. 
This fractional charge is robust against edge manipulations and reflects the properties of the bulk. 
By sampling snapshots of the ground state wave function in the Fock basis, we have demonstrated that the full distribution of the corner charge is peaked around the fractionally quantized value $0{\,}(1/2)$ in the trivial~(topological) phase. 
A similar sampling can be experimentally realized in ultracold atomic settings by using state of the art quantum gas microscopes~\citep{Sherson2010,Bakr2010}; our results pave the way towards the detection of the corner states characterizing HOSPT phases.

\section{Acknowledgments} The authors thank Monika Aidelsburger, Titus Neupert, Maia Vergniory, and Ruben Verresen for stimulating discussions. IL and FP were funded by the European Research Council (ERC) under the European Union’s Horizon 2020 research and innovation program (grant agreement No. 771537). FP acknowledges the support of the DFG Research Unit FOR 1807 through grants no. PO 1370/2- 1, TRR80, and the Deutsche Forschungsgemeinschaft (DFG, German Research Foundation) under Germany’s Excellence Strategy – EXC-2111-390814868. F.G. acknowledges support from the Technical University of Munich - Institute for Advanced Study, funded by the German Excellence Initiative and the European Union FP7 under grant agreement 291763, from the DFG grant No. KN 1254/1-1, and DFG TRR80 (Project F8). YY, FP are supported in part by the National Science Foundation under Grant No.NSF PHY-1748958(KITP) during the Topological Quantum Matter program. 

\bibliography{main.bib}
\section{Appendix}
Here, we briefly show the correspondence between the 2D BBH model and the 2D SL-BHM in the limit of hardcore bosons $U\to\infty$ and a perfect dimerized lattice $t=0$ or $t=1$. The BBH model itself is the 2D version of the SSH model including a $\pi$-flux per plaquette. In the exactly solvable dimerized limits it is sufficient to consider an isolated plaquette, forming a one-dimensional system consisting of four sites with periodic boundary conditions (PBC) for the 2D SL-BHM model and with antiperodic boundary conditions (APBC) for the BBH model. To show the relation we make use of the Jordan-Wigner (JW) transformation. If we denote the creation (annihaltion) operators of the fermions by $\hat{c}^{\dagger}_j{\,}(\hat{c}_j)$ and that of the hardcore bosons by $\hat{a}^{\dagger}_j{\,}(\hat{a}_j)$ then the JW transformation relates them in the following way,
\begin{equation}
\label{JW}
\hat{c}^{\dagger}_{j} = e^{i\pi\sum_{i<j}\hat{n}_i}\hat{a}^{\dagger}_{j},\quad \hat{c}_{j} = \hat{a}_{j}e^{-i\pi\sum_{i<j}\hat{n}_i},
\end{equation}
where $\hat{n}_i = \hat{a}^{\dagger}_i\hat{a}_{i}$. The plaquette Hamiltonians for fermions (F) and hard core bosons (B) are defined by
\begin{align}	\hat{H}^{F}_{P} =& \sum^{3}_{j=1}\left(\hat{c}^{\dagger}_{j}\hat{c}_{j+1} + \text{h.c.}\right)-\left(\hat{c}^{\dagger}_{4}\hat{c}_{1} + \hat{c}^{\dagger}_{1}\hat{c}_{4}\right)\nonumber\\	 \hat{H}^{B}_{P} =& \sum^{4}_{j=1}\left(\hat{a}^{\dagger}_{j}\hat{a}_{j+1} + \text{h.c.}\right),\quad \hat{a}_5\equiv\hat{a}_1,
\end{align}
where $j=\{1,2,3,4\}$ labels the four sites in each plaquette. 
We note that the plaquette Hamiltonian of the BBH model contains a $\pi$-flux per plaquette, since it can be written as
\begin{equation}
\hat{H}^{F}_{P} = \sum^{4}_{j=1}\left(e^{i\phi_{j}}\hat{c}^{\dagger}_{j}\hat{c}_{j+1} + \text{h.c}\right),
\end{equation}
with $\phi_{1}=\phi_{2}=\phi_{3}=0$ and $\phi_4=\pi$.
According to the Peierls substitution the total magnetic flux per plaquette is the sum of all phase factors $\Phi=\sum_{j}\phi_{j}=\pi$. Hence, the APBC of the Hamiltonian $H^{F}_{P}$ accounts for a $\pi$-flux insertion per plaquette. 
As a first step we show the relation excluding the boundary terms, hence $1\leq j<4$,
\begin{align}
\hat{c}^{\dagger}_{j}\hat{c}_{j+1}=&e^{i\pi\sum_{i<j}\hat{n}_i}\hat{a}^{\dagger}_{j}\hat{a}_{j+1}e^{-i\pi\sum_{i<j+1}\hat{n}_i}\nonumber\\
=&\hat{a}^{\dagger}_{j}e^{i\pi\sum_{i<j}\hat{n}_i}e^{-i\pi\sum_{i<j+1}\hat{n}_i}\hat{a}_{j+1}\nonumber\\
=&\hat{a}^{\dagger}_{j}e^{-i\pi\hat{n}_j}\hat{a}_{j+1}\nonumber\\
=&\hat{a}^{\dagger}_{j}\hat{a}_{j+1}
\end{align}
where we used that bosonic operators on different sites commute among each other. Lastly we have to prove that the APBC of the fermionic model give rise to PBC for the bosonic model
\begin{align}
\hat{c}^{\dagger}_{4}\hat{c}_{1}=&e^{i\pi\sum_{i<4}\hat{n}_i}\hat{a}^{\dagger}_{4}\hat{a}_{1}\nonumber\\	=&e^{i\pi\sum^{4}_{i=1}\hat{n}_i}e^{-i\pi\hat{n}_4}\hat{a}^{\dagger}_{4}\hat{a}_{1}\nonumber\\
=&e^{i\pi\hat{N}}e^{-i\pi\hat{n}_4}\hat{a}^{\dagger}_{4}\hat{a}_{1}\nonumber\\
=&-\hat{a}^{\dagger}_{4}\hat{a}_{1}
\end{align}
where we used particle number conservation $\hat{N}=N=2$ (half-filling) and hence $e^{i\pi\hat{N}}=1$. 
Collecting the above results, we find that the BBH model for $t=0$ and $t=1$ can be mapped to 2D SL-BHM by the JW transformation. However, for $t\neq\{0,1\}$ there is no exact mapping even at
$U\to\infty$~\citep{Dubinkin2018}.
\end{document}